\documentclass[pra,twocolumn,superscriptaddress,10pt,noshowpacs]{revtex4}
\usepackage[english]{babel}
\usepackage[T1]{fontenc}
\usepackage[utf8]{inputenc}
\usepackage{graphicx,epstopdf}
\usepackage{amssymb}
\usepackage{amsmath}

\usepackage{amsfonts}
\usepackage{bbm}
\usepackage{color}
\usepackage{latexsym}
\usepackage{caption}
\usepackage{subcaption}
\usepackage{times,txfonts}
\usepackage{color,soul}
\usepackage{listings}
\usepackage{bbold}
\usepackage{comment}

\begin{document}

\title{Non-relativistic Quantum Mechanics on a Twisted Cylindrical Surface}

\author{G. M. Delgado}
\email{gabrieldelgadomelo@alu.ufc.br}
\affiliation{Universidade Federal do Cear\'a, Departamento de F\'{i}sica, 60455-760, Fortaleza, CE, Brazil}

\author{J. E. G. Silva}
\email{euclides@fisica.ufc.br}
\affiliation{Universidade Federal do Cear\'a, Departamento de F\'{i}sica, 60455-760, Fortaleza, CE, Brazil}

\date{\today}

\begin{abstract}
Twisted cylindrical tubes are important model systems for nanostructures, heterostructures, and curved quantum devices. In this work, we investigate the quantum behavior of an electron confined to a twisted cylindrical surface. By first calculating the strain tensor to obtain the induced surface metric, we employ da Costa's formalism to derive the geometry-induced quantum potential. This potential modifies the Schrödinger equation even in the absence of external forces, allowing us to determine the bound states and energy eigenvalues. This was made in the linear and non-linear torsion regime. Furthermore, we analyze two distinct scattering problems: (i) scattering within an infinite cylinder containing a twisted section, and (ii) scattering of a free particle incident upon a finite twisted cylinder. Our goal is to understand how geometry and strain influence the properties of analogous untwisted systems. It turns out that both the linear and non-linear twists yield to a geometric phase into the wave function, while the da Costa potential is kept unchanged. Consequently, the system supports bound states whose energie spectrum is twist independent. For both scattering problems, we find that the transmission probability is insensitive to torsion, whereas it is significantly affected by the particle angular momentum and the cylinder's radius, exhibiting distinct oscillatory behavior. These findings suggest relevant implications for engineering quantum devices based on materials with controlled curvature and twist.
\end{abstract}


\maketitle

\section{Introduction}
\label{sec: introduction}

The influence of geometry on the quantum dynamics of confined systems is a central theme in modern condensed matter physics \cite{bowick2009two,wang2017geometric,WANG201668}. In recent years, two-dimensional (2D) materials such as graphene have attracted significant interest \cite{geim2007rise,kara2012review}, largely because their low-energy electronic excitations behave as massless Dirac fermions, allowing for the study of emergent relativistic effects in tabletop systems~\cite{novoselov2012roadmap}. Furthermore, it is well established that mechanical deformations in low-dimensional structures, such as curvature and torsion in carbon nanotubes~\cite{POPOV200461, pham20222d, Santos_2016}, can be used to decisively modulate their electronic and transport properties. This concept, often termed strain engineering, has been demonstrated experimentally, for instance, through the observation of conductance oscillations in mechanically twisted nanotubes~\cite{cohen2006torsional, pantano2013electronic}.

Among the various theoretical frameworks for describing quantum particles on curved surfaces—such as Dirac's constrained quantization~\cite{LIU2017123} and the geometric momentum approach~\cite{Spittel:15}—the formalism developed by da Costa stands out for its physical clarity~\cite{da1981quantum}. This method effectively confines a particle to a surface by introducing a geometric potential that depends solely on the local mean and Gaussian curvatures, and this formalism can be applied in many geometries \cite{ferrari,catenoid,gomes2020electronic}. This approach is particularly well-suited for investigating the effects of deformations, such as those induced by topological defects like dislocations and disclinations in the crystal lattice~\cite{RevModPhys.80.61}.

In this Letter, we investigate the quantum dynamics of a particle confined to a twisted cylindrical surface. By employing the da Costa formalism with a metric derived from elasticity theory \cite{soutas2012elasticity}, we analyze the system's bound states and scattering properties under both linear and non-linear torsions. We find that the twist induces a geometric phase in the particle's wavefunction but, remarkably, does not alter the quantized energy spectrum. This distinction between phase and energy effects provides insights for engineering quantum states in nanostructures through geometric control.

This paper is organized as follows. In Sec.~\ref{sec: Theoretical Framework}, we establish the theoretical framework by deriving the induced metric for the deformed surface from the Green-Lagrange strain tensor and applying the da Costa formalism to obtain the geometric potential. Sec.~\ref{sec: Linearly deformed cylinder} is dedicated to the study of the linearly deformed cylinder; we solve for the bound states, analyzing the eigenfunctions and eigenenergies, and investigate two distinct scattering scenarios. In Sec.~\ref{sec: Non-Linearly deformed cylinder}, we generalize the analysis to the case of non-linear torsions. Finally, in Sec.~\ref{sec:conclusion}, we present our final remarks and perspectives.
\section{Theoretical Framework}
\label{sec: Theoretical Framework}
\subsection{ Strain Tensor and the Deformed Metric}
\label{subsec:  Strain Tensor and the Deformed Metric}
The mathematical theory of elasticity, developed from the foundational works of Hooke and Mariotte to its modern formulation within continuum mechanics~\cite{soutas2012elasticity}, provides the tools to describe deformed surfaces. A deformation maps a point from an initial position $\vec{r}$ on a surface to a new position $\vec{r'} = \vec{r} + \vec{u}$, where $\vec{u}$ is the displacement field. This alters the squared line element from $ds_0^2 = g_{ij}dq^idq^j$ to $ds^2 = g'_{ij}dq^idq^j$. The relationship between the undeformed metric $g_{ij}$ and the deformed metric $g'_{ij}$ defines the Green-Lagrange strain tensor $\epsilon_{ij}$ as:
\begin{equation}
    g'_{ij} = g_{ij} + 2\epsilon_{ij}.
    \label{eq:metric_strain}
\end{equation}
From this definition, the components of the strain tensor can be derived in terms of the displacement field, resulting in the general expression for the Green-Lagrange strain, which accounts for both linear and non-linear effects of the deformation. The Green-Lagrange strain tensor provides a general description of the deformation of a surface and is given by
\begin{equation}
    \label{eq:general_strain_tensor}
    \epsilon_{ij}=\frac{1}{2}\left(\nabla_i u_j + \nabla_j u_i + \nabla_i u^k \nabla_j u_k\right),
\end{equation}
where $\nabla_i$ denotes the covariant derivative which involves $\Gamma_{il}^k$, the Christoffel  symbols~\cite{carroll2019spacetime}. For a cylindrical surface described by the coordinates $(\phi, z)$, the Christoffel symbols vanish, and the covariant derivative reduces to the partial derivative, i.e., $\nabla_i = \partial_i$.

\begin{figure}[htbp]
    \centering
    \includegraphics[width=0.9\linewidth]{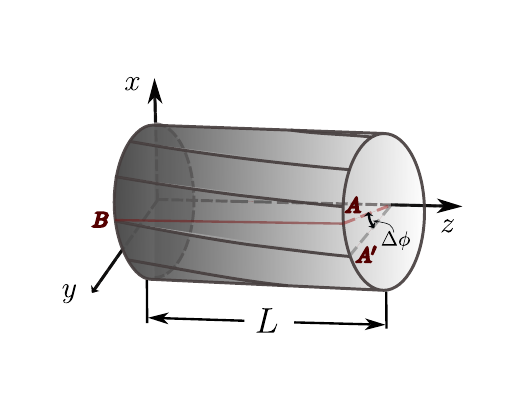}
    \captionsetup{font=small}
    \caption{Schematic of the twisted cylinder model used in this work. The diagram shows the helical deformation of the coordinate lines and the key parameters.}
    \label{fig:torsion_figures}
\end{figure}

For the case of linear torsion, as depicted in Fig.~\ref{fig:torsion_figures}, the contravariant components of the displacement field $\vec{u}$ are
\begin{equation}
	\label{eq:desl_cil_comp}
	u^{\phi} =\Delta\phi=z\alpha, \quad u^r=u^z = 0,
\end{equation}
where $\alpha$ is the constant twist rate (twist per unit length). The corresponding covariant components $u_i$ are obtained by lowering the index with the metric of the undeformed cylinder, $g_{ij} =\begin{pmatrix}
	R^2 & 0 \\
	0   & 1
\end{pmatrix}$. Substituting these quantities into Eq.~(\ref{eq:general_strain_tensor}) yields the strain tensor for the twisted surface:
\begin{equation}
    \label{eq:strain_tensor_twisted_cyl}
	\epsilon_{ij} =
    \begin{pmatrix}
		0 & \frac{1}{2}R^2\alpha \\
		\frac{1}{2}R^2\alpha & \frac{1}{2}R^2\alpha^2
	\end{pmatrix}.
\end{equation}
Consequently, the induced metric on the twisted surface, $g'_{ij}$, is found using the relation $g'_{ij} = g_{ij} + 2\epsilon_{ij}$:
 \begin{equation}
   \label{eq:metric_strain_cyl}
    g_{ij}'=
		\begin{pmatrix}
			R^2 & R^2\alpha \\
			R^2\alpha & 1+R^2\alpha^2
		\end{pmatrix}.
 \end{equation}
Notably, this result can be verified directly by parameterizing the twisted surface with the position vector $\vec{r'}=R\cos(\phi+\alpha z)\hat{i}+R\sin(\phi+\alpha z)\hat{j}+z\hat{k}$ and applying the fundamental definition of the metric, $g'_{ij}=\partial_i\vec{r'}\cdot\partial_j\vec{r'}$.
\subsection{The da Costa Formalism}
\label{subsec:da_costa}
To describe the quantum dynamics on a curved surface, we employ the formalism developed by R. C. T. da Costa~\cite{da1981quantum}. This approach models a particle confined to a surface by considering the limit of a strong binding potential, which effectively reduces the dimensionality of the system from 3D to 2D. The primary result of this formalism is the emergence of a geometric potential in the Schr\"odinger equation for the surface wave function, $\chi_t$. This potential depends solely on the intrinsic and extrinsic curvatures of the surface:
\begin{equation}
    \label{eq:da_costa_potential}
    V_{g} = -\frac{\hbar^2}{2m^*}(M^2 - K),
\end{equation}
where $M$ and $K$ are the Mean and Gaussian curvatures, respectively, which can be expressed in terms of the first ($g_{ij}$) and second ($h_{ij}$) fundamental forms. This method allows for a direct investigation of how the surface geometry impacts the particle's energy spectrum.
For a standard cylinder of radius $R$, the curvatures are $K=0$ and $M=1/(2R)$. Substituting these into Eq.~\eqref{eq:da_costa_potential} yields the well-known geometric potential for a cylinder:
\begin{equation}
    \label{eq:potential_cylinder}
    V_{g}(R) = - \frac{\hbar^2}{8m^*R^2}.
\end{equation}
In eq.(\ref{eq:da_costa_potential}), the term $m^*$ is the \textit{effective mass}, which in general differs from the free particle mass and may depend on position~\cite{silva2021position}.
This approach yields an effective time-dependent Schr\"odinger equation for the surface wave function, $\chi_t$, given by:
\begin{equation}
	\label{eq:Schrodinger_Surface}
	- \frac{\hbar^2}{2m^*} \sum_{i,j=1}^{2} \frac{1}{\sqrt{g}} \frac{\partial}{\partial q_i}
	\left[ \sqrt{g} (g^{-1})_{ij} \frac{\partial \chi_t}{\partial q_j} \right] + V_g\chi_t = i\hbar \frac{\partial \chi_t}{\partial t}.
\end{equation}
This framework allows for a direct investigation of how the geometry of the twisted cylinder impacts the particle's energy spectrum.
\section{Linearly deformed cylinder}
\label{sec: Linearly deformed cylinder}
\subsection{Da Costa Potential of a Twisted Cylinder}
\label{subsec:da_costa_potential_twisted_cylinder}
The da Costa potential is determined by the Mean ($M$) and Gaussian ($K$) curvatures of the surface, which are calculated from the first $g_{ij}$ (Eq.(\ref{eq:metric_strain_cyl})) and second $h_{ij}$ fundamental forms. Where $h_{ij}=\small{
	\begin{pmatrix}
		-R & -R\alpha\\
		-R\alpha & -R\alpha^2
	\end{pmatrix}}$, from these tensors, the curvatures are found to be $K=0$ and $M=1/(2R)$. Substituting these values into Eq.~(\ref{eq:da_costa_potential}) yields the geometric potential for the twisted cylinder:
\begin{equation}
    \label{eq:Vg_twisted_cylinder}
    V_{g}(R) = - \frac{\hbar^2}{8m^*R^2}.
\end{equation}
Notably, this result is identical to the potential of a standard, untwisted cylinder. This implies that a linear torsion, which generates a non-diagonal metric, does not alter the final geometric potential. 
\subsection{Effective Potential and Surface Schr\"odinger Equation}
\label{subsec:effective_potential_and_schrodinger_equation}
\begin{figure*}[htbp]
    \centering

    \begin{subfigure}[t]{0.38\textwidth}
        \centering
        \includegraphics[width=\textwidth]{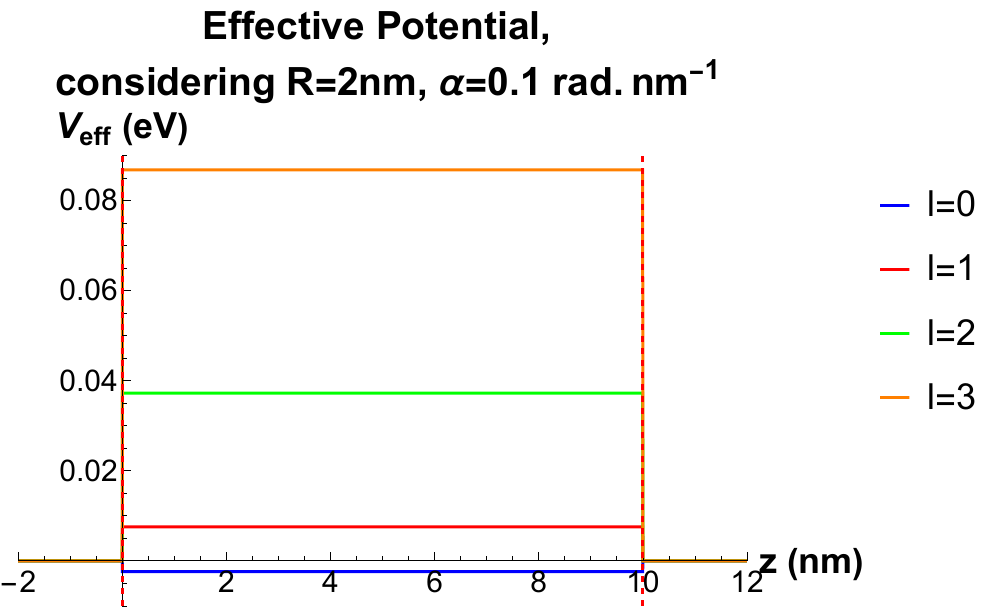}
        \caption{} \label{fig:Veff_a}
    \end{subfigure}
    \hspace{5em}
    \begin{subfigure}[t]{0.42\textwidth}
        \centering
        \includegraphics[width=\textwidth]{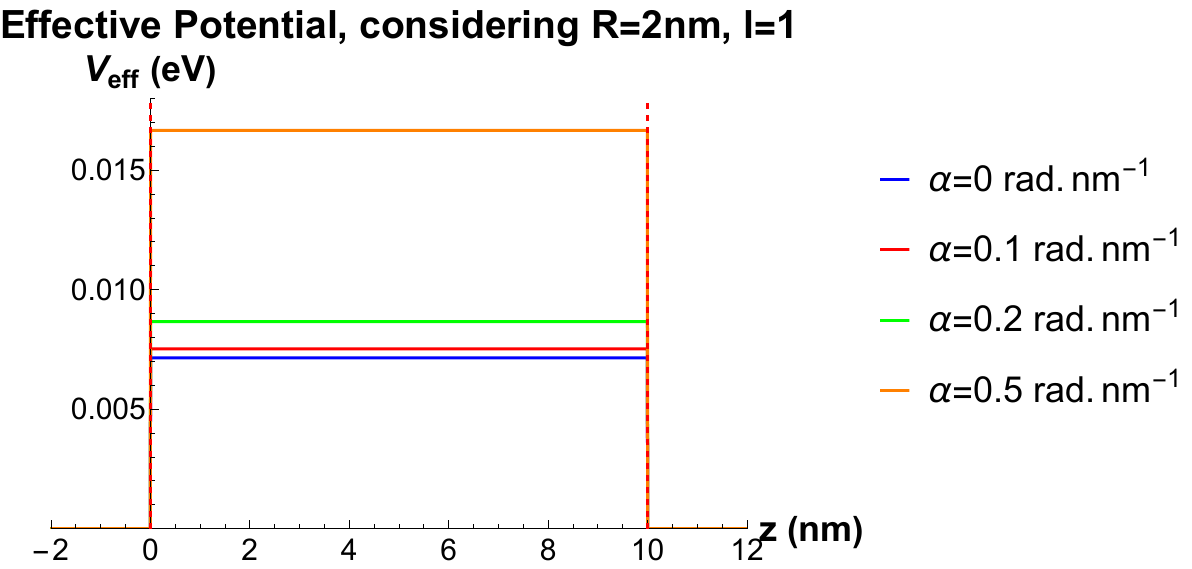}
        \caption{} \label{fig:Veff_b}
    \end{subfigure}
    
    \captionsetup{font=small}
    \caption{Analysis of the effective potential ($V_{\text{eff}}$), assuming the free electron mass ($m^*\approx m_e$). 
    \textbf{(a, b)} $V_{\text{eff}}$ as a function of the longitudinal coordinate $z$ for different values of $l$ and $\alpha$.}
    \label{fig:Veff_complete}
    \end{figure*}

To obtain the time-independent Schr\"odinger equation for the twisted cylinder, we begin by expanding the Laplace-Beltrami operator in Eq.~(\ref{eq:Schrodinger_Surface}). This requires the inverse of the deformed metric from Eq.~(\ref{eq:metric_strain_cyl}), which is
\begin{equation}
    \label{eq:inv_metric_strain_cyl}
    g^{ij} = (g^{-1})_{ij} = \frac{1}{R^2}
    \begin{pmatrix}
        1+R^2\alpha^2 & -R^2\alpha \\
        -R^2\alpha   & R^2
    \end{pmatrix}.
\end{equation}
After performing the derivatives and simplifying, the stationary Schr\"odinger equation on the surface becomes:
\begin{equation}
    \label{eq:EDP_phi_z}
    -\frac{\hbar^2}{2m^*}\left( g^{\phi\phi}\frac{\partial^2\psi}{\partial \phi^2} + g^{zz}\frac{\partial^2\psi}{\partial z^2} + 2g^{\phi z}\frac{\partial^2\psi}{\partial\phi\partial z} \right) + V_g(R)\psi = \varepsilon \psi.
\end{equation}
Given the azimuthal symmetry of the problem, the variables can be separated by proposing a solution of the form 
\begin{equation}
\psi(\phi, z) = Z(z)e^{il\phi},    
\end{equation}
where $l$ is an integer angular momentum quantum number. This reduces the partial differential equation~(\ref{eq:EDP_phi_z}) to a one-dimensional ordinary differential equation for $Z(z)$:
\begin{equation}
	\label{eq:EDO_z_covariant}
		- \frac{\hbar^2g_{\phi\phi}}{2m^*R^2}\frac{d^2Z}{dz^2}+il\frac{\hbar^2g_{\phi z}}{m^*R^2}\frac{dZ}{dz}+\left(V_g(R)+\frac{\hbar^2g_{zz}}{2m^*R^2}l^2\right)Z=\varepsilon Z.
\end{equation}
Equation (\ref{eq:EDO_z_covariant}) describes the longitudinal dynamics of the particle under an effective potential, $V_{\text{eff}}$, which includes the centrifugal term. The effective potential for a given mode $l$ is given by
\begin{equation}
    \label{eq:V_eff_final_covariant}
    V_{\text{eff}} = V_g(R) + \frac{\hbar^2 g_{zz}}{2m^*R^2}l^2 = \frac{\hbar^2}{2m^*R^2}\left( g_{zz}l^2-\frac{1}{4} \right),
\end{equation}
where $l=0,\pm1,\pm2, \dots$ and we have substituted the explicit form for $V_g(R)$. Note that $g_{zz} = 1+R^2\alpha^2$ contains the dependence on the torsion. The plots of the effective potential in terms of different parameters are shown in Fig.(\ref{fig:Veff_complete}). To make these graphs, we take an approximation of the effective mass approximately equal to the mass of the free electron ($m^*\approx m_e$).
\subsection{Eigenfunctions and Eigenenergies}
\label{subsec:Eigenfunctions and eigenenergies}
The longitudinal wave equation Eq.~(\ref{eq:EDO_z_covariant}) has a general solution of the form 
\begin{equation}
Z(z)=Ae^{r_1z}+Be^{r_2z},    
\end{equation}
where the roots $r_{1,2}$ are given by
\begin{equation}
\label{roots}
r_{1,2}=il\alpha \mp \sqrt{\frac{2m^*}{\hbar^2}(V_{\text{eff}}-\varepsilon)-l^2\alpha^2}.   
\end{equation}
It is worthwhile to mention that the twist constant $\alpha$ does not appear in the squared root together with the effective potential $V_{\text{eff}}$ in Eq.\ref{roots}, as expected. In order to interpret this result, let us perform the following change upon the wave function of form
\begin{equation}
\label{changeofwavefunction}
    Z(z)=e^{i\alpha l} u(z)
\end{equation}
Substituting Eq.(\ref{changeofwavefunction}) into Eq.(\ref{eq:EDO_z_covariant}) leads to the wave equation for $u$ in the form
\begin{equation}
\label{eq:edo_z_u}
- \frac{\hbar^2}{2m^*}\frac{d^2 u}{dz^2}+V_{\text{eff}}^*(R)u=\varepsilon u,
\end{equation}
where the new effective potential $V_{\text{eff}}^*(R)$ is purely real and independent of the torsion:
\begin{equation}
\label{eq:effective_linear_pot_*}
V_{\text{eff}}^*(R)=V_g+\frac{\hbar^2}{2m^*}\frac{l^2}{R^2} = \frac{\hbar^2}{2m^*R^2}\left(l^2-\frac{1}{4} \right).
\end{equation}
The change on the wave function performed in Eq.(\ref{changeofwavefunction}) can be understood as an unitary transformation $Z=U(\alpha){u}$, where the unitary operator $U(\alpha)$ is given by
\begin{equation}
    U(\alpha)=e^{i\alpha l}.
\end{equation}
Since the wave functions $Z(z)$ and $u(z)$ are related by an unitary transformation, they have the same energy spectrum and eingenstates. Thus, the twist produces a kind of geometric phase, similar to ones found in curved nanostructures as nanocones \cite{cone} and M\"{o}bius ribbons \cite{mobius}.


\vspace{1em}
\noindent\textit{Case 1}: $\varepsilon > V_{\text{eff}}^*$ (\textit{Bound States})

In this regime, Eq.~(\ref{eq:edo_z_u}) can be written as $u''=-k^2u$, with a real wavevector $k=\sqrt{\frac{2m^*}{\hbar^2}\left(\varepsilon- V_{\text{eff}}^*\right)}$. The general solution for the full longitudinal wavefunction $Z(z)$ is therefore:
\begin{equation}
\label{eq:Z(z)_general_sol}
Z(z)=e^{il\alpha z}\left(A\sin{(kz)}+B\cos{(kz)}\right).
\end{equation}
Applying the confining boundary conditions $Z(0)=Z(L)=0$, results in $Z_{nl}(z)=A_{nl}\sin \left(\frac{n\pi z}{L}\right)e^{il\alpha z}$. The normalized wavefunctions are:
\begin{equation}
\label{eq:autofuncoes_espacial}
\begin{aligned}
\psi_{nl}(\phi,z)&=\frac{1}{\sqrt{\pi RL}}\sin\left(\frac{n\pi z}{L}\right)e^{il(\phi+\alpha z)},
\end{aligned}
\end{equation}
where $n=1, 2, 3, \dots$ and $l=0, \pm 1, \pm 2, \dots$ are the longitudinal and azimuthal quantum numbers, respectively. A key feature of these wavefunctions is the torsion-induced phase factor, $e^{il\alpha z}$. However, this phase vanishes when calculating the surface probability density, $|\psi_{nl}|^2$, which is therefore independent of the torsion parameter $\alpha$.
The corresponding eigenenergies are:
	\begin{equation}
      \begin{aligned}
            \label{eq:autoenergias}
			\varepsilon_{nl}(R) &=\frac{\hbar^2k_n^2}{2m^*}+ V_{\text{eff}}^*,\\
            &=\frac{\hbar^2}{2m^*}\frac{n^2\pi^2}{L^2}+\frac{\hbar^2}{2m^*R^2}\left(l^2-\frac{1}{4}\right).
      \end{aligned}
	\end{equation}
Notably, and perhaps counter-intuitively, the energy spectrum for the bound states is independent of the torsion parameter $\alpha$. This indicates that a linear torsion, while modifying the phase of the wavefunction, does not shift the quantized energy levels of the system.  
In the limit of an infinitely wide cylinder ($R\to\infty$), the energy reduces to that of an infinite quantum well, $\varepsilon_{n} \to \frac{\hbar^2}{2m^*}\frac{n^2\pi^2}{L^2}$.
The other case is the infinite wire of radius R ($L\rightarrow\infty$), the energy reduces to $\varepsilon_{nl}\rightarrow\frac{\hbar^2}{2m^*R^2}\left(l^2-\frac{1}{4}\right)$
The ground state of the system corresponds to $(n,l)=(1,0)$.

\vspace{0.75em}
\noindent\textit{Case 2}: $\varepsilon \leq V_{\text{eff}}^*$ (\textit{Non-oscillatory Solutions})

In this regime, the general solution for $Z(z)$ is a non-oscillatory function of the form:
\begin{equation}
Z(z) = e^{il\alpha z} \left( c_1 e^{\kappa z} + c_2 e^{-\kappa z} \right),
\end{equation}
where $\kappa=\sqrt{\frac{2m^*}{\hbar^2}\left(V_{\text{eff}}^*-\varepsilon\right)}$ is real. Applying the boundary condition $Z(0)=0$ requires $c_2 = -c_1$, which simplifies the solution to $Z(z) = 2c_1 e^{il\alpha z} \sinh(\kappa z)$. The second condition, $Z(L)=0$, then demands that $\sinh(\kappa L)=0$. Since $L>0$, this is only possible if $\kappa=0$, which in turn corresponds to the trivial solution $Z(z)=0$. Therefore, no physically acceptable bound states exist in this energy regime.

\subsection{Scattering by a Finite Twisted Cylinder Section}
\label{subsec: Scattering by a Finite Twisted Cylinder Section}
\begin{figure}[htbp]
    \centering
    \begin{subfigure}[b]{0.6\linewidth}
        \centering
        \includegraphics[width=\textwidth]{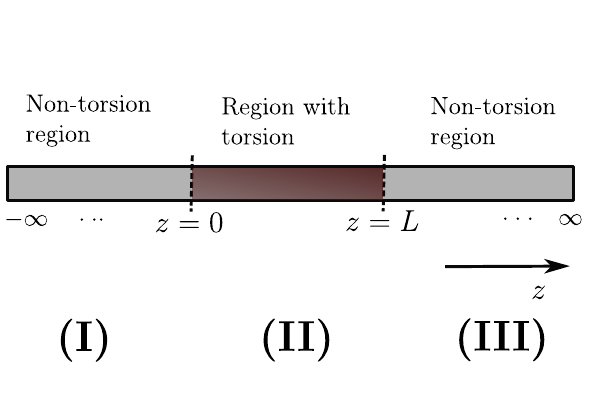}
        \caption{}
        \label{fig:scattering_setup}
    \end{subfigure}
    \hfill
    \begin{subfigure}[b]{0.6\linewidth}
        \centering
        \includegraphics[width=\textwidth]{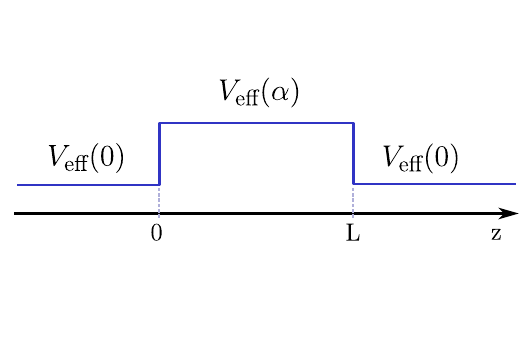}
        \caption{}
        \label{fig:potential_scattering}
    \end{subfigure}
    
    \captionsetup{font=small}
    \caption{Illustration of the scattering problem. 
    \textbf{(a)} Schematic of the physical setup, where a finite twisted cylinder section (Region II) of length $L$ is embedded in an infinite untwisted cylinder (Regions I and III). 
    \textbf{(b)} The corresponding effective potential profile $V_{\text{eff}}$ encountered by a particle with a given angular momentum mode. The torsion creates a potential step which acts as a scattering center.}
    \label{fig:scattering_figures}
\end{figure}

We now analyze the scattering of a quantum particle by a finite linearly twisted section of an infinite cylinder. As illustrated in Fig.~\ref{fig:scattering_figures}, a twisted section of length $L$ (Region II) is embedded within an infinite untwisted cylinder, acting as a scattering center. The torsion induces an effective potential step between the regions, with a height given by $\Delta V_{\text{eff}} = V_{\text{eff}}(R,\alpha) - V_{\text{eff}}(R,0) = \hbar^2 l^2 \alpha^2 / (2m^*)$.
To determine the reflection ($r$) and transmission ($t$) coefficients, we solve the time-independent Schrödinger equation in each region. The longitudinal wavefunction $Z(z)$ is described by propagating plane waves in the untwisted regions (I and III) and by a superposition of modes with complex wavevectors, $r_1$ and $r_2$, in the twisted region (II):
\begin{equation}
	\label{eq:scattering_solutions}
	Z(z) = 
	\begin{cases}
		e^{ikz}+re^{-ikz}, & \text{for } z < 0 \text{ (Region I)} \\
		Ae^{r_1 z}+Be^{r_2 z}, & \text{for } 0 \leq z \leq L \text{ (Region II)} \\
		te^{ikz}, & \text{for } z > L \text{ (Region III)}.
	\end{cases}
\end{equation}
The coefficients are found by imposing boundary conditions at the interfaces $z=0$ and $z=L$, which require the continuity of the wavefunction and the probability current.
\begin{equation}
	\label{eq:continuity_psi}
	Z_{(I)}(0)=Z_{(II)}(0) \quad \text{and} \quad Z_{(II)}(L)=Z_{(III)}(L).
\end{equation}
The boundary conditions for the wavefunction derivatives are determined by ensuring the continuity of the probability current, $j$, at the interfaces. In the twisted region (II), the current contains an additional torsion-dependent term: $j=\frac{i\hbar}{2m^*}\left(\psi\frac{\partial\psi^*}{\partial z}-\psi^*\frac{\partial\psi}{\partial z}\right)-\frac{\hbar}{m^*}l\alpha|\psi|^2$. Due to this term, the standard condition of continuous derivatives ($Z'$) is no longer valid. Instead, imposing the continuity of the current itself at the boundaries ($j_{(I)}(0)=j_{(II)}(0)$ and $j_{(II)}(L)=j_{(III)}(L)$) yields the correct relations for the derivatives:
\begin{equation}
\label{eq:derivatives_conditions}
\begin{cases}
Z'_{(I)}(0)=Z'_{(II)}(0)-il\alpha Z_{(II)}(0),
\\[1em]
Z'_{(III)}(L)=Z'_{(II)}(L)-il\alpha Z_{(II)}(L).
\end{cases}
\end{equation}
These conditions, along with wavefunction continuity from Eq.~(\ref{eq:continuity_psi}), form the system of four linear equations required to solve the scattering problem.
\begin{equation}
\label{eq:system_scattering}
\begin{cases}
1+r=A+B,
\\
ik(1-r)=r_1A+r_2B-il\alpha (A+B),
\\
te^{ikL}=Ae^{r_1L}+Be^{r_2L},
\\
ikte^{ikL}=r_1Ae^{r_1L}+r_2Be^{r_2L}-il\alpha (Ae^{r_1L}+Be^{r_2L}).
\end{cases}
\end{equation}
Solving the system of equations for the scattering coefficients yields the transmission ($T$) and reflection ($R$) probabilities. Our numerical analysis reveals a "surprising" result: for all energies above the propagation threshold ($E > V_{\text{eff}}(0)$), the twisted section is effectively \textit{transparent}, with $T=1$ and $R=0$. This perfect transmission occurs regardless of the torsion parameter $\alpha$. This result can be understood by recalling that the effective potential governing the wavefunction's amplitude, Eq.~(\ref{eq:effective_linear_pot_*}), is identical in both the twisted and untwisted regions, thus eliminating any potential step for the propagating part of the wave.

\begin{figure}[htbp]
    \centering
    \begin{subfigure}[b]{0.85\linewidth}
        \centering
        \includegraphics[width=\textwidth]{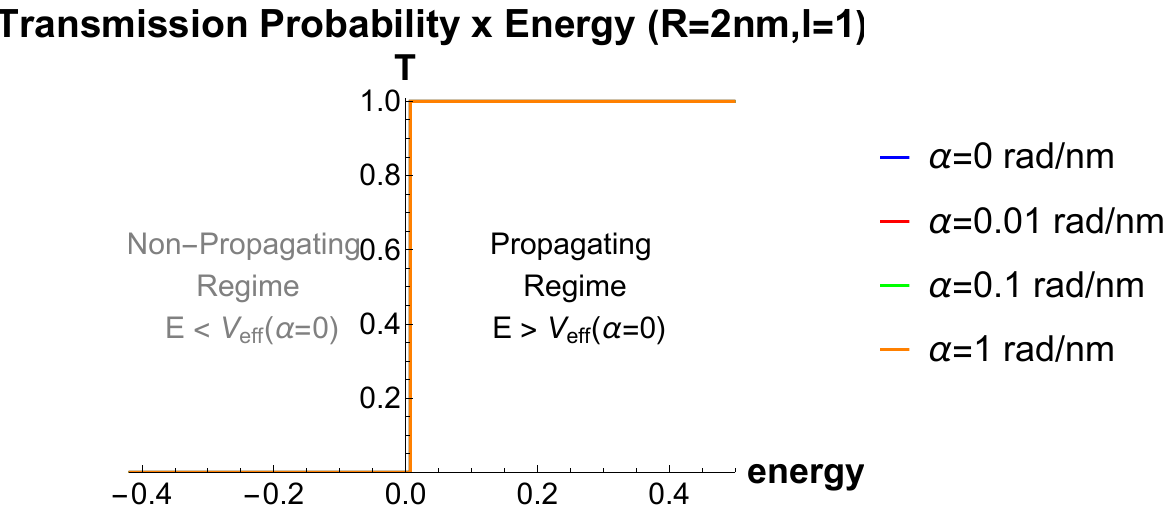}
        \caption{}
        \label{fig:scattering_Trans_alpha}
    \end{subfigure}
    
    \begin{subfigure}[b]{0.85\linewidth}
        \centering
        \includegraphics[width=\textwidth]{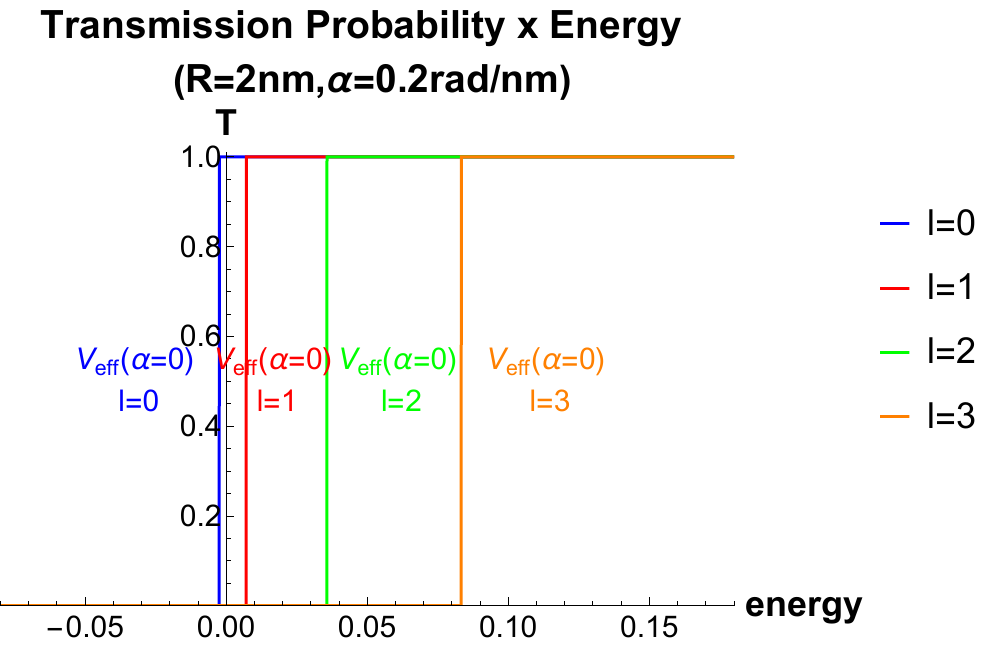}
        \caption{}
        \label{fig:scattering_Trans_l}
    \end{subfigure}
    
    \begin{subfigure}[b]{0.85\linewidth}
        \centering
        \includegraphics[width=\textwidth]{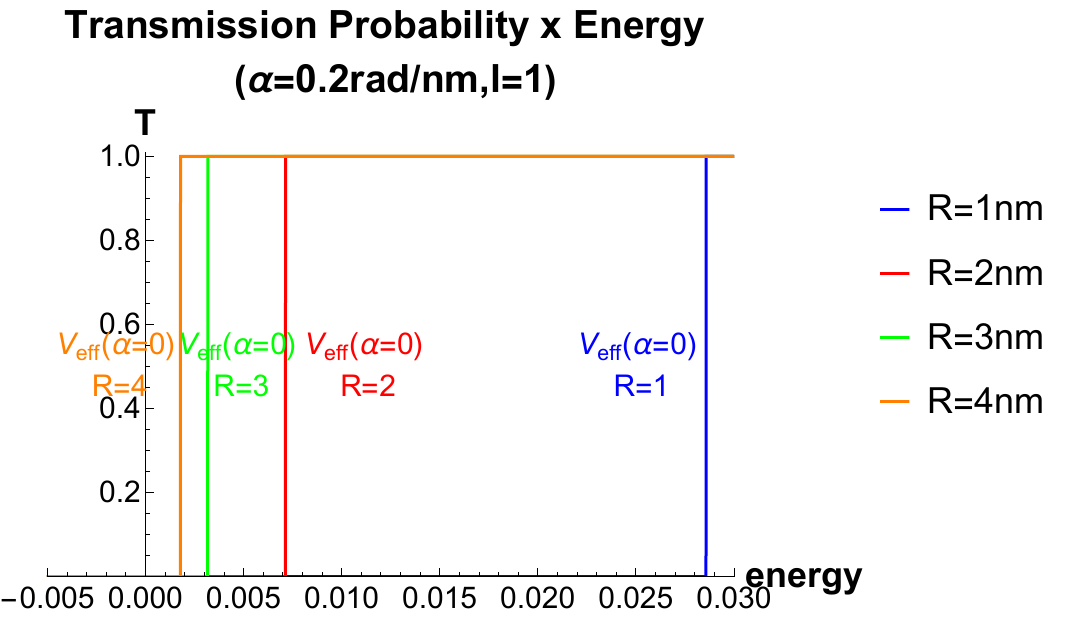}
        \caption{}
        \label{fig:scattering_Trans_R}
    \end{subfigure}
    
    \captionsetup{font=small}
    \caption{Transmission probability ($T$) versus incident energy ($E$). \textbf{(a)}~The transmission is independent of the torsion parameter $\alpha$. \textbf{(b)}~The energy threshold for transmission increases for higher angular momentum modes $l$. \textbf{(c)}~The energy threshold decreases for a larger cylinder    radius $R$. Fixed parameters are indicated in the plots.}
    \label{fig:scattering_Trans_complete}
\end{figure}

This behavior is illustrated in Fig.~\ref{fig:scattering_Trans_complete}, where the transmission exhibits a step-function profile. While the transmission is always perfect above the threshold, the position of this energy threshold is strongly dependent on the system's geometry and quantum numbers. As shown in panels (b) and (c), the threshold \textit{increases with higher angular momentum $l$} and \textit{decreases with a larger cylinder radius $R$}, confirming that these parameters, unlike the torsion, effectively control the onset of propagation.
\subsection{Scattering of a Free Particle by a Twisted Cylinder}
\label{subsec:Scattering of a Free Particle by a Twisted Cylinder}

\begin{figure}[htbp]
    \centering
    \includegraphics[width=0.5\columnwidth]{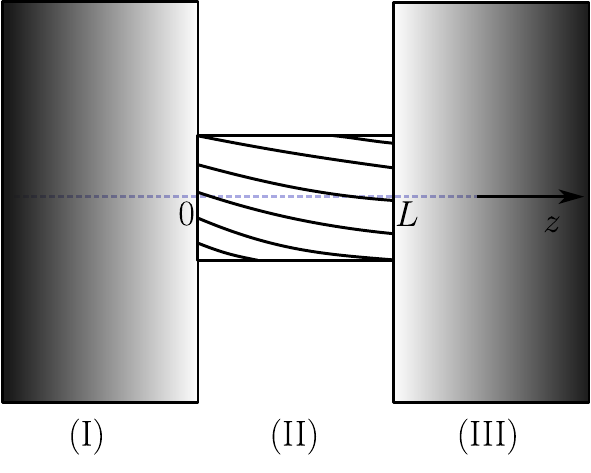}
    \captionsetup{font=small}
    \caption{Illustration of the scattering problem. Schematic of the physical setup, where a finite twisted cylinder section (Region II) of length $L$ is embedded in an infinite free space (Regions I and III).}
    \label{fig:free_particle_scattering_setup}
\end{figure}

We now adapt the formalism from the previous section to a different physical scenario: the scattering of a free particle by a finite twisted cylinder. The setup, illustrated in Fig.~(\ref{fig:free_particle_scattering_setup}), consists of a twisted cylinder section (Region II) of length $L$ embedded between two regions of free space (Regions I and III). As the particle moving in the $z$ direction enters Region II, it encounters an effective potential step, $\Delta V_{\text{eff}} = \frac{\hbar^2}{2m^*R^2}\left( \left(1+R^2\alpha^2 \right)l^2-\frac{1}{4} \right).$
To adapt the model for a free incident particle, the wavevector in the untwisted regions is redefined as $k=\sqrt{2m_e\varepsilon/\hbar^2}$. By applying the approximation $m^*\approx m_e$, the resulting system of equations becomes formally identical to the previous case. However, the new definition of $k$ yields distinct physical results. The twisted cylinder now acts as a scattering barrier with non-perfect transmission ($T<1$), and our numerical analysis confirms that probability is conserved ($R+T=1$).

\begin{figure}[htbp]
    \centering
    \begin{subfigure}[b]{0.9\linewidth}
        \centering
        \includegraphics[width=\textwidth]{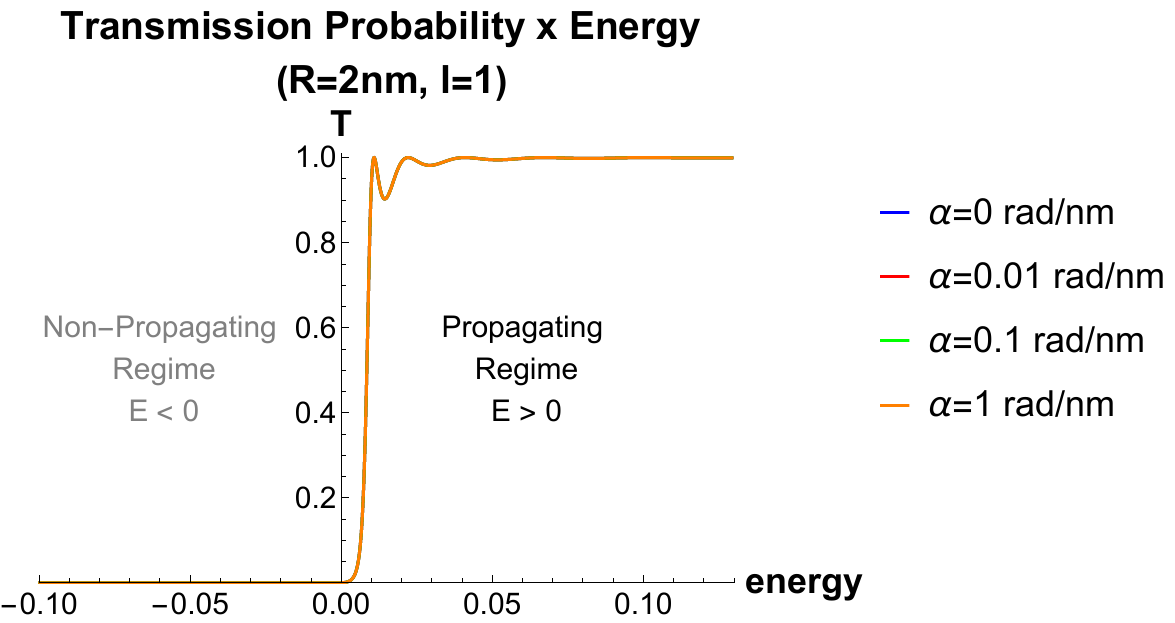}
        \caption{}
        \label{fig:scattering_Trans_free_alpha}
    \end{subfigure}
    
    \begin{subfigure}[b]{0.8\linewidth}
        \centering
        \includegraphics[width=\textwidth]{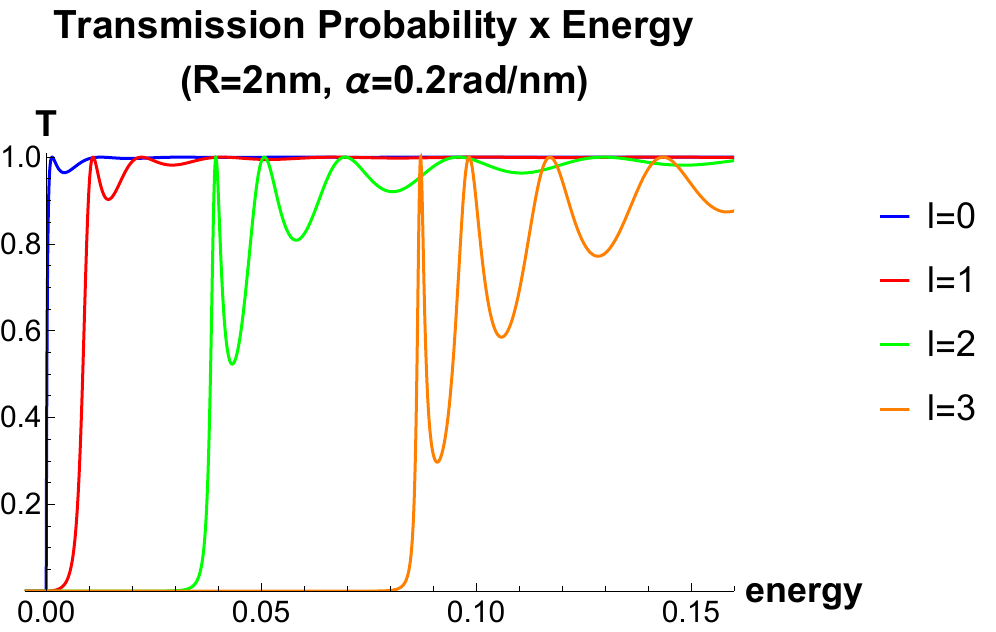}
        \caption{}
        \label{fig:scattering_Trans__free_l}
    \end{subfigure}
    
    \begin{subfigure}[b]{0.8\linewidth}
        \centering
        \includegraphics[width=\textwidth]{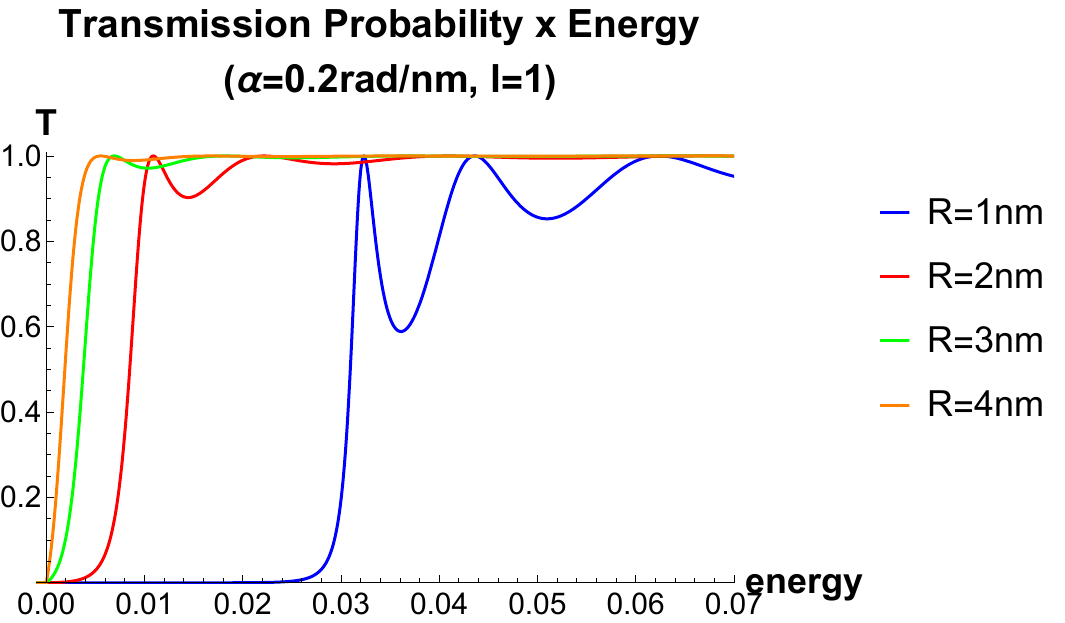}
        \caption{}
        \label{fig:scattering_Trans_free_R}
    \end{subfigure}
    \captionsetup{font=small}
    \caption{Transmission probability ($T$) versus incident energy ($E$). \textbf{(a)}~The transmission spectrum is insensitive to the torsion parameter $\alpha$. \textbf{(b)}~The spectrum exhibits oscillations that depend strongly on the angular momentum $l$. \textbf{(c)}~Similarly, the transmission is strongly modulated by the cylinder radius $R$.}
    \label{fig:scattering_Trans_free_complete}
\end{figure}

The calculated transmission probabilities are shown in Fig.~(\ref{fig:scattering_Trans_free_complete}). The results reveal several key features: the spectrum is largely \textit{insensitive to the torsion parameter $\alpha$}, consistent with the fact that the effective potential barrier can be reduced to a torsion-independent form. In contrast, the transmission is \textit{strongly modulated by the angular momentum $l$ and the cylinder radius $R$}. Increasing $l$ raises the energy threshold for transmission, while increasing $R$ lowers it. For energies above this threshold, the transmission exhibits \textit{oscillatory behavior}.
\section{Non-linearly Deformed Cylinder}
\label{sec: Non-Linearly deformed cylinder}

In the previous section, the torsion parameter $\alpha$ was taken as a constant. We now generalize this problem by considering a non-uniform torsion where the twist rate depends on the longitudinal coordinate, $\alpha = \alpha(z)$, while keeping the cylinder radius constant, $R = \text{const}$.

\subsection{The Deformed Metric and Da Costa Potential}
\label{subsec: Deformed Metric and Da Costa Potential}

For this scenario, we calculate the metric tensor from its fundamental definition, $g_{ij}=\partial_i\vec{r}\cdot\partial_j\vec{r}$, using the position vector for the non-linearly twisted surface: $\vec{r}=R\cos\left(\phi+\alpha(z)z \right)\hat{i}+R\sin\left(\phi+\alpha(z)z \right)\hat{j}+z\hat{k}$.
This calculation yields
\begin{equation}
   \label{eq:metric_nonLin_cyl}
    g_{ij}=
		\begin{pmatrix}
			R^2 & R^2f(z) \\
			R^2f(z) & 1+R^2f(z)^2
		\end{pmatrix},
\end{equation}
where the function $f(z)$ is $f(z)=\alpha(z)+z\alpha'(z)$. Notably, if $\alpha$ is constant, then $f(z)=\alpha$, and this metric correctly reduces to the one found for the linear case in Eq.~(\ref{eq:metric_strain_cyl}). The inverse of this metric is given by:
\begin{equation}
    \label{eq:inv_metric_nonLin_cyl}
    g^{ij} =
    \begin{pmatrix}
        \frac{1}{R^2}+f(z)^2 & -f(z) \\
        -f(z)  & 1
    \end{pmatrix}.
\end{equation}
For the non-linearly twisted cylinder, we find that the curvatures are $K=0$ and $M=1/(2R)$, respectively. Substituting these into Eq.~(\ref{eq:da_costa_potential}) yields the already obtained geometric potential Eq.(\ref{eq:potential_cylinder},\ref{eq:Vg_twisted_cylinder}).
Remarkably, this result is identical to the potential of a standard, untwisted cylinder. This implies that the geometric potential is insensitive to torsion, even in the non-linear regime.
\subsection{Effective Potential and Surface Schrödinger Equation}
\label{subsec:Effective Potential and Surface Schr_dinger Equation for Non Linear Case}
Substituting the inverse metric Eq.~(\ref{eq:inv_metric_nonLin_cyl}) into the surface Schrödinger equation (Eq.~(\ref{eq:Schrodinger_Surface})) and separating variables leads to the following ODE for the longitudinal component $Z(z)$:
\begin{equation}
	\label{eq:EDO_z_nonLin}
	\begin{split}
		- \frac{\hbar^2}{2m^*} \frac{d^2Z}{dz^2} &+ ilf(z)\frac{\hbar^2}{m^*}\frac{dZ}{dz}+ \\
		&+ \left[ V_g(R) + \frac{\hbar^2}{2m^*} \left( f(z)^2 + \frac{1}{R^2} \right)l^2 + il\frac{\hbar^2}{2m^*}f'(z) \right] Z = \varepsilon Z.
	\end{split}
\end{equation}
Unlike the linear case, simply defining the term in the square brackets as an effective potential is problematic because it includes an imaginary, z-dependent term. To obtain a physically meaningful potential for bound states, we follow the procedure from Sec.~(\ref{subsec:Eigenfunctions and eigenenergies}) and redefine the wavefunction as $Z(z)=\zeta(z)u(z)$.
\begin{equation}
  \begin{split}
       \label{eq:edo_z_zeta_u_nonlin}
		- \frac{\hbar^2}{2m^*}\frac{\zeta'' u}{\zeta}- \frac{\hbar^2}{2m^*}u''- \frac{\hbar^2}{m^*}\frac{\zeta' u'}{\zeta}+il\frac{\hbar^2}{m^*}f(z)\frac{\zeta' u}{\zeta}+il\frac{\hbar^2}{m^*}f(z) u'+& \\ 
        +\left[ V_g(R) + \frac{\hbar^2}{2m^*} \left( f(z)^2 + \frac{1}{R^2} \right)l^2 + il\frac{\hbar^2}{2m^*}f'(z) \right]u=\varepsilon u.
  \end{split}
\end{equation}
By choosing the phase function $\zeta(z)$ such that $\zeta'(z)/\zeta(z)=ilf(z)$, which implies $\zeta(z) \propto \exp\left(il\int_{0}^{z}f(\xi)d\xi\right)$, the ODE for the amplitude function $u(z)$ simplifies into one in the same form as 
Eq.~(\ref{eq:edo_z_u}) from the linear torsion analysis, indicating that the final effective potential and the governing equation for the wavefunction's amplitude are invariant under non-linear torsion. 

\vspace{1em}
\noindent\textit{Case 1}: $\varepsilon > V_{\text{eff}}$ (\textit{Bound States})

In this regime, the general solution for the longitudinal wavefunction $Z(z)$ is:
\begin{equation}
\label{eq:Z(z)_general_sol_nonlin}
Z(z)=\exp\left(il\int_{0}^{z}f(\xi)d\xi\right)\left(A\sin{(kz)}+B\cos{(kz)}\right),
\end{equation}
where $k=\sqrt{\frac{2m^*}{\hbar^2}\left(\varepsilon - V_{\text{eff}}\right)}$ and $V_{\text{eff}} = V_g+\frac{\hbar^2 l^2}{2m^*R^2}$. Applying the confining boundary conditions, $Z(0)=Z(L)=0$ results in $Z_{nl}(z)=A_{nl}\sin \left(\frac{n\pi z}{L}\right)\exp\left(il\int_{0}^{z}f(\xi)d\xi\right)$, so, the normalized wavefunction is:
\begin{equation}
\label{eq:autofuncoes_espacial_nonlin}
\begin{aligned}
\psi_{nl}(\phi,z)&=\frac{1}{\sqrt{\pi RL}}\sin\left(\frac{n\pi z}{L}\right)\exp\left[il\left(\phi+\int_{0}^{z}f(\xi)d\xi\right)\right],
\end{aligned}
\end{equation}
where $f(z)=\alpha(z)+z\alpha'(z)$, and $n, l$ are the longitudinal and azimuthal quantum numbers. A key feature is the torsion-induced phase factor. If $\alpha=\text{const.}$, the phase integral reduces to $il\alpha z$, correctly recovering the linear case results from Eq.~(\ref{eq:autofuncoes_espacial}).
The corresponding eigenenergies of Eq.(\ref{eq:autofuncoes_espacial_nonlin}) are identical to the linear torsion case, meaning that for a constant radius, the energy spectrum is completely insensitive to the profile of the torsion.

\vspace{0.75em}
\noindent\textit{Case 2}: $\varepsilon \leq V_{\text{eff}}$ (\textit{Non-oscillatory Solutions})

Following the same procedure as in the linear case, applying the boundary conditions $Z(0)=Z(L)=0$ to the non-oscillatory solutions leads to the trivial solution $Z(z)=0$. Therefore, no bound states exist in this energy regime.
\section{Final Remarks and Perspectives}
\label{sec:conclusion}
We investigated the quantum mechanics of a particle on a twisted cylindrical surface, analyzing both bound and scattering states. Our central finding is that linear and specific non-linear torsions act as a geometric gauge field: they introduce a significant z-dependent phase into the eigenfunctions but \textit{remarkably leave the bound-state energy spectrum and surface probability density invariant}. This subtlety persists in the scattering regime, where we found the transmission probability to be \textit{largely insensitive to the torsion parameter $\alpha$}, while being strongly modulated by the cylinder's radius ($R$) and the particle's angular momentum ($l$).

These results, which distinguish between geometric phase effects and energy effects, open several avenues for future research. Natural extensions include investigating more general deformations where the radius also varies, $\alpha(z)$ and $R(z)$, and applying the formalism to the Dirac equation to study relativistic quasiparticles. Furthermore, the calculated energy spectrum can be used to derive thermodynamic observables, following standard methods found in textbooks such as Ashcroft and Mermin~\cite{ashcroft1976solid}.

\section*{Acknowledgments}
I would like to thank Professor J. E. G. Silva for fruitful discussions and valuable insights. G. M. Delgado acknowledges financial support from the Coordena{\c c}{\~a}o de Aperfei{\c c}amento de Pessoal de N{\'i}vel Superior (CAPES), Brazil - Finance Code 001.

\bibliographystyle{unsrt}
\bibliography{references}

\end{document}